# Erbium-excess gallium garnets

Chen Yang[1,2], Haozhe Wang[3], Lun Jin[1], Xianghan Xu[1], Danrui Ni[1], Jeff D. Thompson[2], Weiwei Xie[3] and R. J. Cava[1]

[1] Department of Chemistry, Princeton University, Princeton, New Jersey, 08544

[2] Department of Electrical and Computer Engineering, Princeton University, Princeton, New Jersey, 08544

[3] Department of Chemistry, Michigan State University, East Lansing, Michigan, 48824

*Abstract*

A series of garnets of formula $Er_{3+x}Ga_{5-x}O_{12}$ is described, for which we report the crystal structures for both polycrystalline and single-crystal samples. The $x$ limit in the garnet phase is between 0.5 and 0.6 under our conditions, with the Er fully occupying the normal garnet site plus half-occupying the octahedral site at $x = 0.5$ in place of the Ga normally present. Long-range antiferromagnetic order with spin ice-like frustration is suggested by the transition temperature ($T_N \approx 0.8K$) being much lower than the Curie-Weiss theta. The magnetic ordering temperature does not depend on the Er excess, but there is increasing residual entropy as the Er excess is increased, highlighting the potential for unusual magnetic behavior in this system.

**Introduction**

Garnets have a long-standing history of being used in laser technology and as substrates in a range of applications ([1–5]). Of particular interest to us are garnets containing Erbium (Er), due to their unique properties, which include long spin coherence times and narrow spectral linewidths. The doping of yttrium iron garnet ($Y_3Fe_5O_{12}$) single crystals with Er is explored as a potential proof-of-concept system for transduction ([6,7]). These findings suggest that Er-doped garnets may offer a possible avenue for the development of materials for quantum computing. To optimize the quality of the transduction process required for the transmission of quantum information over long distances, it is important to understand the formula of Er garnets and if possible to vary the concentration of Er ions in the garnet structure. The site occupancy flexibility of Ga, Al, and Fe ions thus makes rare earth gallium garnets, rare earth aluminum garnets, and rare earth iron garnets particularly interesting materials to study in this context.

Structurally, garnets can be written with the formula $R_3(M'_{tet})_3(M_{oct})_2O_{12}$, where the R site has 8 oxygen neighbors, $M_{oct}$ has 6 oxygen neighbors in octahedral geometry and $M'_{tet}$ has 4 oxygen neighbors in a tetrahedral geometry. Rare earth excess may therefore be possible in the garnet structure for rare-earth ions where an octahedral coordination environment may not be unfavorable [7]. In this vein, a non-stoichiometric gadolinium gallium garnet was first reported in 1973, in a study of the changing lattice parameter for different gallium oxide contents [8] However, the absence of a crystal structure determination and chemical formula in that report limit its usefulness. By gaining a better understanding of the crystal structure and chemical composition of a non-stoichiometric rare earth excess garnet, we hope to optimize their materials quality and pave the way for the development of more advanced quantum computing materials. We note that the rare earth ion in the gallium garnet is the only magnetic ion present for our materials (as opposed to the Fe case) and that we have found that Er excess is not possible under our conditions for the Er-aluminate garnet.

In addition, these materials are of interest due to their lattice geometry, which may yield geometrically frustrated magnetism (Figure 1). Geometrically frustrated magnetism has been a topic of significant interest in condensed matter physics [9–14], recently, in particular, in the investigation of spin-liquid behavior [9]. Recently, the hyperkagome structures formed by the rare earth sites in garnets have been found to cause magnetic frustration in some cases [11,12] and the corner-sharing triangular lattices of magnetic rare earths in some garnets have been shown to lead to spin-liquid behavior, as confirmed by muon and neutron spectroscopies [13–15].

In this paper, we focus on the role of additional rare earth ions in a series of rare-earth-excess gallium garnets, of formula $Er_{3+x}Ga_{5-x}O_{12}$, up to the solubility limit of about $x = 0.5$ and investigate the resulting crystal structure and magnetic properties. Our findings indicate that excess Er (i.e. $x$ greater than 0) is possible to obtain in this material under normal synthetic conditions. The refined crystal structure shows that the normally occupied Er site remains filled in the Er-excess garnet and that the excess Er is accommodated in the octahedral sites. Our results further suggest that although the temperature-dependent magnetic susceptibility (per Er) above 1.8 K is not significantly impacted by the excess Erbium, there is a systematic increase in the residual entropy with increasing Er excess.

**Experimental**

*Polycrystalline Sample Synthesis:* Polycrystalline samples of rare earth excess erbium gallium garnet, $Er_{3+x}Ga_{5-x}O_{12}$, were synthesized using a conventional high-temperature ceramic method. The starting materials $Er_2O_3$ (Thermo scientific 99.99%) and $Ga_2O_3$ (Sigma Aldrich 99.99%) were carefully measured in molar ratios and ground together. The resulting mixture was then transferred into an alumina crucible, which was heated in a muffle furnace, in air, at 1500 °C for 14 days with intermittent grinding. The heating and cooling rates were both set to 3 °C/min to ensure a controlled and uniform reaction process. In order to assess the advancement of the chemical reaction, powder X-ray diffraction (PXRD) measurements were carried out after each grinding step utilizing a Brucker D8 FOCUS diffractometer with Cu Kα radiation ($\lambda_{K\alpha}$ = 1.5406 Å). Upon the observation of a single phase, PXRD data were collected for subsequent LeBail and Rietveld refinements (Figure S1). The data obtained were analyzed by means of the TOPAS and GSAS-II software packages to determine the crystal structure and fine-tune the atomic positions in the synthesized materials.

*Single Crystal Synthesis:* A crystal containing both $Er_{3+x}Ga_{5-x}O_{12}$ and $ErMgGaO_4$ was grown using the optical floating zone method. The starting material $Er_2O_3$ was dried overnight at 900 °C. MgO and $Ga_2O_3$ were dried overnight at 400 °C. A total mass of 15g was weighed while remaining dry in a molar ratio of 2:1:2 ($Er_2O_3$: MgO: $Ga_2O_3$). The powder was mixed well using a mechanical mortar and pestle. The powder was compressed into long rods using balloons and a hydraulic press. The crystal was grown in the air at atmospheric pressure with a growth rate of 0.5mm/h. An airflow of 2L/min was used. The feed and seed shafts were rotated at 15rpm in opposite directions to enable better heat convection inside the molten zone.

A fragment with dimensions 0.145 × 0.077 × 0.065 mm³ was picked up, mounted on a nylon loop with paratone oil, and its diffraction measured using a XtalLAB Synergy, Dualflex, Hypix single crystal X-ray diffractometer with an Oxford Cryosystems low-temperature device, operating at T = 300.0(2) K. CRI Data were measured using ω scans using Mo K α radiation ( λ = 0.71073 Å, micro-focus sealed X-ray tube, 50 kV, 1 mA). The total number of runs and images was based on the strategy calculation from the program CrysAlisPro 1.171.42.79a (Rigaku OD, 2022). Data reduction was performed with correction for Lorentz polarization. The integration of the data using a cubic unit cell yielded a total of 26749 reflections to a maximum θ angle of 40.55° (0.55 Å resolution), of which 504 were independent (average redundancy 53.073, completeness = 99.0%, $R_{int}$ = 4.82%). A numerical absorption correction based on gaussian integration over a multifaceted crystal model was applied using spherical harmonics, implemented in the SCALE3 ABSPACK scaling algorithm. The $Er_{3.2}Ga_{4.8}O_{12}$ crystal structure was determined by subtracting the weak diffraction pattern of $ErMgGaO_4$ from the observed pattern (Figure S2).

*Magnetic Property Measurements:* Magnetic measurements on single phase polycrystalline samples were performed using a Quantum Design Dynacool Physical Property Measurement System (PPMS) equipped with a vibrating sample magnetometer (VSM). The magnetic susceptibility ($\chi$) was defined as the ratio of magnetization (M) to the applied field (H). Zero-field cooled (ZFC) magnetic data were acquired under a magnetic field of H = 0.1 T (1000 Oe) over a temperature range spanning from 1.8 K to 300 K. A modified Curie-Weiss law $\chi - \chi_0 = C/(T - \theta_{CW})$ was fitted to the inverse $\chi$ data over the temperature range of 150-250 K in order to derive the Curie constant (C) and Weiss temperature ($\theta_{CW}$), with $\chi_0$ representing a temperature-independent correction factor. The effective magnetic moment ($\mu_{eff}$) per formula unit, expressed in Bohr magnetons ($\mu_B$), was calculated from the constant C in the above expression using $\sqrt{8C}$. Field-dependent magnetometry measurements were collected within at temperatures of 1.8 K and 300 K under magnetic fields ranging from -9 T to +9 T (-90.000 to + 90,000 Oe).

*Specific Heat Capacity Measurements:* Heat capacity measurements of the stoichiometric and Er-excess samples were carried out using a PPMS equipped with a Helium-3 (He-3) system. Both addenda and sample heat capacity data were collected in the absence of an external magnetic field, within a temperature range spanning from 0.5 K to 10 K. The addenda were carefully chosen to provide an appropriate reference for the sample, and their heat capacity contributions were subtracted from the total heat capacity data to obtain the true heat capacity of the samples. The measurements were repeated multiple times to ensure the reproducibility and reliability of the results.

**Results and Discussion**

Single crystal X-ray diffraction was used to determine the crystal structure of the material with excess Er ions in the garnet lattice. The Er-excess garnet was found to crystallize in a cubic lattice with the space group I*a*-3*d*, which is the same space group as is found for the stoichiometric material. The excess $Er^{3+}$ ions were found to preferentially occupy the octahedral sites in the garnet structure, as postulated would be possible, in a random fashion. The tetrahedral sites were found to be too small to accommodate the rare-earth ions (Figure 1).

The results of the powder X-ray diffraction characterization confirm and extend this result. Polycrystalline samples between the stoichiometric garnet and $x = 0.5$ in the Er-excess garnet were straightforwardly synthesized under our conditions. Analysis of the power diffraction data shows that the limit of Er excess lies between $x = 0.5$ and $x = 0.6$ for $Er_{3+x}Ga_{5-x}O_{12}$, demonstrated by the appearance of diffraction peaks from $Er_3GaO_6$ by $x = 0.6$. The cubic lattice parameter was found to increase as the Erbium excess increased. The PXRD patterns for all samples were refined using LeBail fitting to determine the lattice parameter, and the normalized lattice parameter $a/a_0$ versus Er concentration $x$ is plotted in Figure 2. A nearly linear increase in lattice parameters with increasing Er content is observed, although for $x = 0.4$ the peaks in the diffraction pattern were unusually broad. The evolution in peak position around $2\theta = 17.5$ degrees, the (220) peak, is a good way to assess the sample quality (inset, Figure 2a).

Antiferromagnetic coupling without long-range ordering is observed through the magnetic characterization of these materials at temperatures above 1.8 K. Curie-Weiss fitting was applied to the magnetic susceptibility data, where the behavior is given by $\chi(T) = C/(T - \theta_{CW})$, (where C is the Curie constant and $\theta_{CW}$ is the Curie-Weiss temperature.) The fitting showed that the stoichiometric and the most Er excess samples have highly similar magnetic behavior, as their Curie Weiss thetas and effective moments per Er are within experimental error of each other (Figure 3a). The negative $\theta_{CW}$ values obtained from the fitting indicate that for both the stoichiometric and Er excess samples the magnetic moments are relatively strongly coupled for (rare earth ions) antiferromagnetically. No long-range magnetic ordering was observed above 1.8 K, which, along with the low temperature change in slope of the inverse chi plot indicates that there is magnetic frustration in these samples. The field-dependent Magnetization data (Figure 3b) show that both the stoichiometric and Er-excess material behave as simple paramagnets at 300 K and that there is a significant ferromagnetic contribution to their susceptibility by 2 K, consistent with expectations for frustrated magnets and with previous results in the stoichiometric case ([11,15]).

Heat capacity data were collected to access temperatures below 1.8K on polycrystalline pellets. The peak near 0.8 K (Figure 4a) is a signature for a magnetically ordered system. It is noteworthy that the transition temperature for all samples is almost the same despite the varying Er concentrations. Curiously, the 0.8 K heat capacity peak decreases in magnitude as the Erbium excess is increased (Figure 4a). Thus the Er excess, which is seen only to structurally impact the octahedral site, significantly suppresses the magnetic ordering of the normally present garnet Er site.

The total heat capacity consists of three parts for a magnetic material, which are $C_{tot} = C_{el} + C_{ph} + C_{mag}$. The phonon contribution, $C_{ph}$, has been approximated and subtracted through the characterization of the non-magnetic garnet analog $Lu_3Ga_5O_{12}$. At low temperatures, $C_{el} = \gamma T$ dominates, while $C_{ph} = \beta T^3$. Given that garnets are insulators, $C_{el}$ can be approximated as zero. In this case, an approximation of $C_{tot} = C_{mag}$ can be made. The integration of $C_{mag}/T$ was directly performed on the data set to obtain the $\Delta S_{mag}$ curve. The stoichiometric sample's $\Delta S_{mag}$ is within error of Rln2 at ~ 3.7 K, indicating that the system is Ising-like and fully magnetically ordered (Figure 4b). However, residual entropy remains for all the non-stoichiometric samples, which suggests that not all Er spins are aligned at temperatures down to 0.5 K. The implication is that as the Er-excess concentration increases, more triangular substructures are created between the Er spins on the normal garnet and the partially occupied octahedral sites, resulting in increased magnetic frustration and spin-liquid behavior. This appears to be similar to what is observed in "stuffed spin ice" but it would have to be supported through further characterization.

**Conclusions**

This paper presents the chemical formula and properties of rare earth excess gallium garnets $Er_{3+x}Ga_{5-x}O_{12}$ ($Er_3Ga_3Ga_{2-x}Er_xO_{12}$ structurally) for the first time. Polycrystalline samples up to $x=0.5$ were synthesized by the conventional solid-state method. Single crystal diffraction data was also refined and reported for the first time for a rare earth excess garnet, $Er_{3.2}Ga_{4.8}O_{12}$. The temperature-dependent magnetic susceptibilities and the heat capacities of the Er-excess garnet samples reveal interesting magnetic behavior. No long-range magnetic ordering was observed between 1.8 and 300 K in the magnetic susceptibility measurements. Heat capacity data showed that the ordering temperature, $T_N$, remained around 0.8K for all samples; while the residual entropy increased with increasing Er content, deviating from the simple Ising magnet behavior seen for the stoichiometric substance. The frustrated magnetism is deduced to come from hyper-pyrochlore-lattice-derived geometric frustration.

**Acknowledgments**

This material is based upon work supported by the U.S. Department of Energy, Office of Science, National Quantum Information Science Research Centers, Co-design Center for Quantum Advantage (C2QA) under contract number DE-SC0012704. The research at Michigan State University was funded by the U.S. Department of Energy, Basic Energy Sciences under contract number DE-SC0023568.

**Figures**

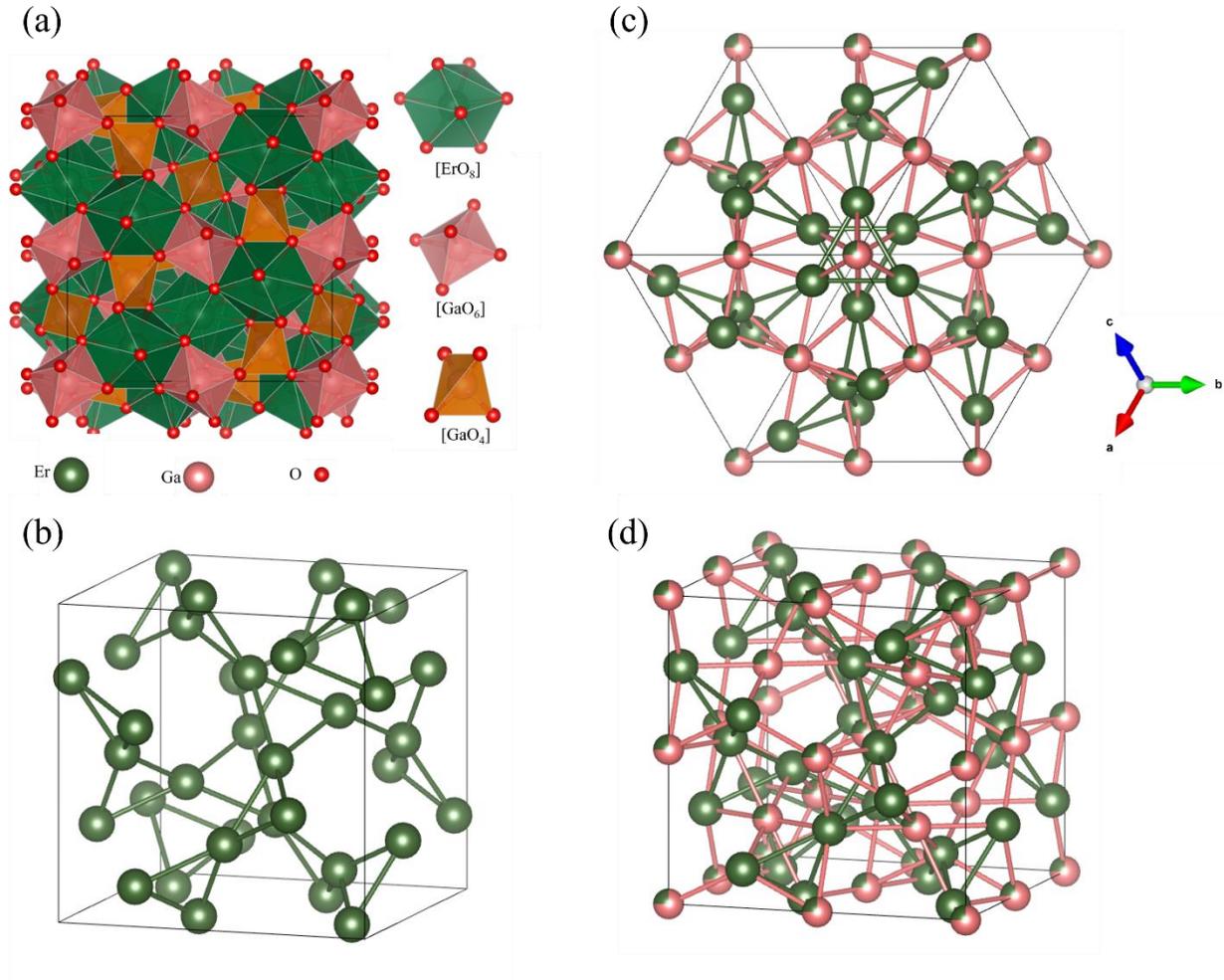

**Figure 1. The crystal structure of Er$_3$Ga$_5$O$_{12}$ and the Er-excess material.** (a) The polyhedral representation (b) The Er hyper-kagome net in Er$_3$Ga$_5$O$_{12}$. (c) The Er hyper-pyrochlore net in Er$_{3+x}$Ga$_{5-x}$O$_{12}$ looking down from [111]. The green bonds link two Er dodecahedral sites. The pink bonds link dodecahedral Er and octahedral Er sites. The green spheres in (b) and (c) are the Er3+ and Ga3+ ions respectively. (d) The Er plus octahedral Er/Ga hyper-pyrochlore triangular net in the Er excess material Er$_{3.2}$Ga$_{4.8}$O$_{12}$.

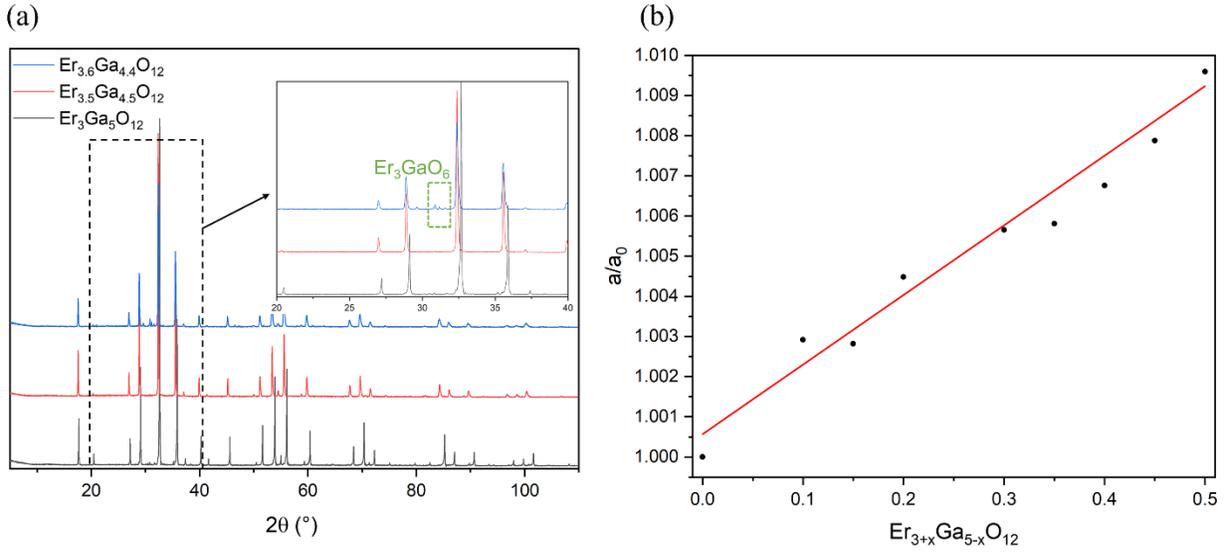

**Figure 2**. **Powder X-ray diffraction characterization of the Er-Ga garnets** (a) patterns of Er-excess $Er_3Ga_5O_{12}$ and an undoped sample. (See Figure S1 for the GSASII refinement of $Er_3Er_{0.5}Ga_{4.5}Ga_3O_{12}$.) Inset: The low angle area of the diffraction patterns zoomed in, showing the evidence for the solubility limit of Er in this garnet. (b) The normalized lattice parameter change with increasing Erbium excess in the $Er_{3+x}Ga_{5-x}O_{12}$ garnet follows Vegard's Law.

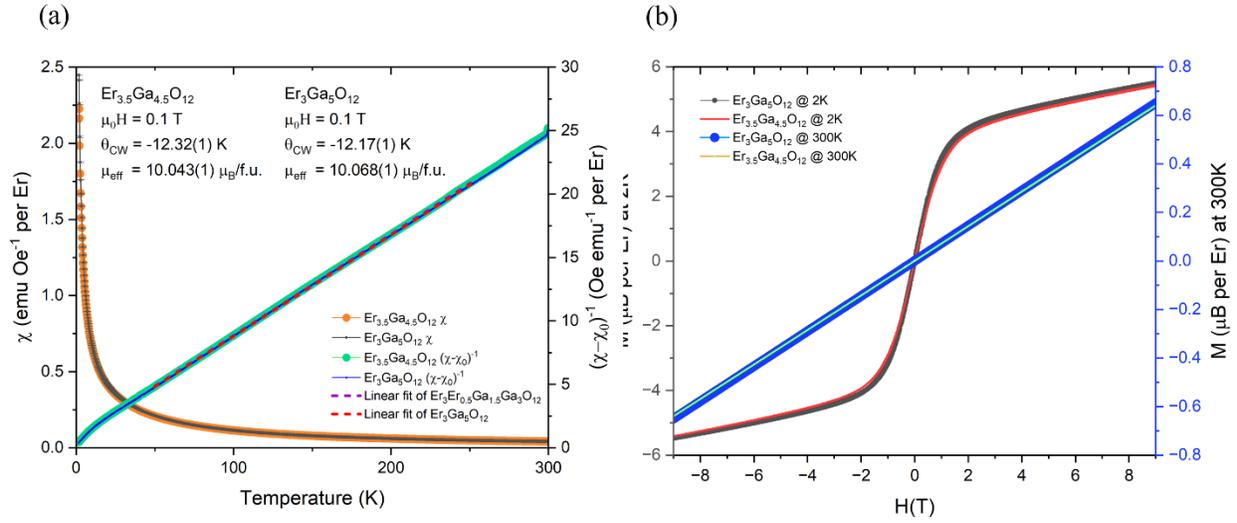

**Figure 3**. **Magnetic characterization of the materials** (a) The magnetic susceptibility of $Er_3Ga_5O_{12}$ with and without Er exces measured at 0.1 T and fitted with the Curie-Weiss law. (b) Magnetization data of $Er_3Ga_5O_{12}$ and the Er excess material at 300 K and 2K measured in magnetic fields from -9 T to +9 T.

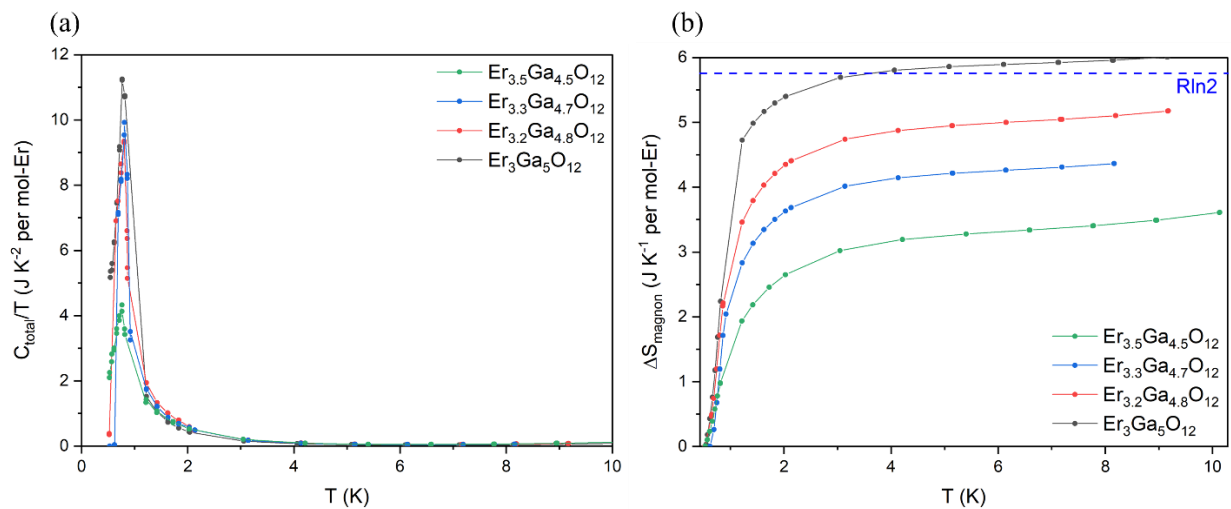

**Figure 4**. **The specific heat capacity characterization**. (a) Heat capacity per Er of both the stoichiometric and Er excess garnets measured down to 0.5 K. (b) The Integrated Entropy of $Er_3Ga_5O_{12}$ plus various Er excess compositions.